\documentclass{ifacconf}

\usepackage{graphicx}      
\usepackage{natbib}        
\usepackage{comment}
\usepackage{amsmath} 
\usepackage{amssymb}  
\usepackage{algorithm}
\usepackage{multicol}
\usepackage{multirow}
\usepackage{bm}
\usepackage{booktabs}
\usepackage{subcaption}

\newenvironment{proof}{\par\noindent{\bf Proof.}}{\hfill$\blacksquare$\par}

\let\classAND\AND
\let\AND\relax
\usepackage{algorithmic}

\let\AND\classAND
\AtBeginEnvironment{algorithmic}{\let\AND\algoAND}

\floatname{algorithm}{Algorithm}

\begin{document}
\begin{frontmatter}

\title{Discrete-time Two-Layered Forgetting RLS Identification under Finite Excitation\thanksref{footnoteinfo}} 

\thanks[footnoteinfo]{This work was supported by JSPS Grant-in-Aid for JSPS Fellows Grant Number JP23KJ1923.}

\author[First]{Satoshi Tsuruhara} 
\author[Second]{Kazuhisa Ito} 

\address[First]{Graduate School of Engineering and Science, Functional Control Systems, Shibaura Institute of Technology, JSPS Research Fellow, $307$ Fukasaku, Minuma, Saitama $3378570$, Japan\\(e-mail: nb23110@shibaura-it.ac.jp).}
\address[Second]{Department of Machinery and Control Systems, Shibaura Institute of Technology, 307 Fukasaku, Minuma, Saitama 3378570, Japan\\
(e-mail: kazu-ito@shibaura-it.ac.jp).}

\begin{abstract}                
In recent years, adaptive identification methods that can achieve the true value convergence of parameters without requiring persistent excitation (PE) have been widely studied, and concurrent learning has been intensively studied.
However, the parameter convergence rate is limited for the gradient-based method owing to small parameter update gain, and even the introduction of forgetting factors does not work sufficiently.
To address this problem, this study proposes a novel discrete-time recursive least squares method under finite excitation (FE) conditions using two forgetting factors (inner and outer) and an augmented regressor matrix comprising a sum of regressor vectors.
The proposed method ensures the PE condition of the augmented regressor matrix under FE conditions of the regressor vector and allows the properly design of the forgetting factor without estimator windup and/or destabilization of the system.
Numerical simulations demonstrate its effectiveness by comparing it with several conventional methods.
\end{abstract}


\end{frontmatter}
\section{INTRODUCTION}
A recursive least squares (RLS) algorithm is significantly practical owing to its fast parameter convergence rate. This has been widely applied, particularly in combination with exponential forgetting (EF), to improve robustness against characteristic changes in plants (\cite{landou,goodwin}). However, if the persistent excitation (PE) condition is not satisfied, the algorithm cannot guarantee to achieve true value convergence. Furthermore, under non-PE conditions, an estimator wind-up of the covariance matrix occurs, which significantly reduces its robustness to noise.
This implies that the forgetting factor cannot be designed arbitrarily. To avoid this problem, several forgetting factor algorithms, such as directional forgetting (DF) and exponential resetting (ER), which can guarantee the existence of the upper and lower bounds of the covariance matrix, have been proposed (\cite{DF,FF}); however, the DF cannot achieve true value convergence even when the PE condition is satisfied. Unlike the EF, the ER has difficulty setting the threshold value for resetting and a slow convergence rate.

In contrast, several algorithms have been studied in recent years to achieve true value convergence of parameters without PE conditions (\cite{CL, Composite, DREM}). Solving this problem is a significant challenge in adaptive identification and control, because the PE condition is too strict to be satisfied in practical applications. Among them, concurrent learning adaptive identification extends the gradient method by adding a modified term using recorded data to achieve true value convergence of the parameters under the finite excitation (FE) condition in which the parameters should be excited at a finite time period, which relaxes the PE condition. Various extensions have been proposed, including the integral (\cite{Int}) and discrete-time (\cite{DCL}) types.
However, because it is based on a normalized gradient-type method, its parameter convergence rate is considerably lower than that of the RLS algorithm, thereby preventing its widespread application.
To address this problem, finite-time CL (\cite{FT}) and DF-based CL algorithms (\cite{DF-CL}) have been proposed. However, the parameter update gain cannot be designed sufficiently large to guarantee stability, and thus, the transient performance has not been effectively improved.

In this study, based on the background above, we propose a novel discrete-time RLS algorithm with a fast convergence rate and high robustness under FE conditions. In recent years, several important results have been proposed for the RLS algorithm without requiring a PE condition (\cite{RLS1, RLS2}). In (\cite{RLS1}), the regularization term was appropriately incorporated. In (\cite{RLS2}), the regressor vector and parameter update law were modified to guarantee the positive definiteness of the information matrix to achieve zero convergence of the parameter error. Although these methods are effective, they switch the covariance matrix updating with a time step, and the designer must set their thresholds, which significantly affect their performance. In particular, (\cite{RLS2}) is for continuous system design and applies the EF algorithm to regressor updates; it is difficult to achieve sufficient performance if appropriate covariance matrix updates cannot be implemented.

In this paper, inspired by the aforementioned studies, we propose a two-layer forgetting factor algorithm for discrete-time systems comprising an inner and outer forgetting factor. In an inner loop, the proposed method updates the regressor matrix based on the regressor vector using the DF algorithm with an inner forgetting factor.
In a outer loop, on the other hand, the proposed method constructs an RLS algorithm using the EF algorithm as the outer forgetting factor with the augmented regressor matrix.
The proposed method can easily satisfies the PE condition for the augmented regressor matrix, thereby guaranteeing the existence of upper and lower bounds of the covariance matrix, even under the FE condition. Thus, the value of the forgetting factor of EF can be properly designed, in other words, it can be designed close to zero, under FE conditions, which has the strong advantage of allowing the explicit tuning of the convergence rate of the parameters in the design phase to avoid an estimator windup. In addition, the condition number is improved, which improves the robustness against noise and/or characteristic changes. The existence of upper and lower bounds of the covariance matrix of the proposed method was demonstrated, and the global exponential stability of the parameter errors was proved. 
The proposed method is expected to serve as a safe, robust, and versatile discrete-time identification approach that ensures stability despite its simplicity.
Numerical simulations demonstrate the effectiveness of the proposed method.
\section{Problem Formulation and Preliminaries}
Consider the following linear parametric model:
\begin{align}
y(k+1) = \phi^T(k)\theta,
\end{align}
where $y\in\mathbb{R}$, $\phi\in\mathbb{R}^{n}$, and $\theta\in\mathbb{R}^{n}$ denote the output, regressor vector, and unknown parameter vector for time step $k\geq 0$, respectively.

\noindent \textbf{\underline{Problem Formulation}} This paper aims to propose discrete-time parameter identification algorithm with exponentially fast convergence rate under FE condition, thus, $\lim_{k\to\infty}\|\hat{\theta}(k)-\theta\|\to 0$ exponentially by depending on forgetting factor, where $\hat{\theta}(k)\in\mathbb{R}^n$ denotes the estimated parameter vector.

As mathematical preliminaries, the following two definitions are known for regressor by (\cite{tao}) and (\cite{Slotine}):
\begin{defn}\label{def:PE}
The bounded matrix signal $\Phi(k)\in\mathbb{R}^{n\times n}$, $n\geq 1$, is PE if there exist $\delta>0$ and $\alpha_{\phi}>0$ such that
\begin{align}
\sum_{k=\sigma}^{\sigma+\delta}\Phi(k)\Phi^T(k)\geq\alpha_{\Phi}I,\ \forall \sigma\geq k_0.
\end{align}
\end{defn}

\begin{defn}\label{def:FE}
The bounded matrix signal, $\Phi(k)\in\mathbb{R}^{n\times n}$, $n\geq 1$, is exciting over time sequence set $\{\sigma_0,\sigma_0+1,\ldots,\sigma_0+\delta_\Phi\},\ \delta_\Phi>0,\sigma_0\geq k_0$, if for some $\alpha_\Phi>0$ it holds that
\begin{align}
\sum_{k=\sigma_0}^{\sigma_0+\delta_\Phi}\Phi(k)\Phi^T(k)\geq \alpha_{\Phi}I
\end{align}
This condition is known as FE.
\end{defn}



\section{Proposed methods}
In this section, to relax PE condition, the augmented regressor matrix $\Omega(k)\in\mathbb{R}^{n\times n}$ is derived from a regressor vector $\phi(k)$ using the DF algorithm. 
Correspondingly, an auxiliary vector $M(k)\in\mathbb{R}^n$ is introduced and an augmented identification error $\Omega(k)\hat{\theta}(k)-M(k)$ is defined, which includes past data unlike a conventional identification error with only instantaneous data.
Then, the EF-RLS algorithm is designed based on these signals.

\subsection{Extention to augmented regressor matrix}
The augmented regressor matrix $\Omega(k)$ and the auxiliary vector $M(k)$ are made up of the sum of the normalized regressor vectors $\phi(k)/m(k)$, that is, $\Omega(k+1)=\Omega(k)+{\phi(k)\phi^T(k)}/{m^2(k)}$ and $M(k+1)=M(k)+{\phi(k)y(k+1)}/{m^2(k)}$, where $m(k)=\sqrt{1+\phi^T(k)\phi(k)}$ in Algorithm 1.
By continuing to accumulate normalized regressor vectors, the upper bounds of the augmented regressor matrix cannot be guaranteed. Therefore, in this study, we prevented this problem by introducing a DF algorithm that decomposes the matrix into the directions of the excited and unexcited signals and forgets only the matrix in the excited direction.
Let $\mu\in[0,1)$ be the value of the forgetting factor, call it inner forgetting factor.
Note that, unlike the general forgetting factor, the closer it is to $1$, the higher the forgetting performance, and $0$ corresponds to no forgetting.
These algorithms are summarized in Algorithm \ref{alg:DF-regressor}.

\begin{algorithm}
\caption{Generation of augmented regressor matrix based on DF inspired by (\cite{DF-CL})}
\begin{algorithmic}\label{alg:DF-regressor}
\STATE{\textbf{Step 1}: Set $\Omega(0)=0_{n\times n}$, $M(0)=0_{n\times 1}$
\STATE{\textbf{Step 2}: \IF{$\mathrm{rank}
(\Omega (k))<\mathrm{rank}\left(\Omega (k)+\frac{\phi(k)\phi^T(k)}{m^2(k)}\right)$}
\STATE{$\Omega(k+1)=\Omega(k)+\frac{\phi(k)\phi^T(k)}{m^2(k)}$}
\STATE{$M(k+1)=M(k)+\frac{\phi(k)y(k+1)}{m^2(k)}$}
\ELSE
\STATE{$\Omega(k+1)=\Omega(k)-\mu\frac{\Omega(k)\phi(k)\phi^T(k)}{\phi^T(k)\Omega(k)\phi(k)}\Omega(k)+\frac{\phi(k)\phi^T(k)}{m^2(k)}$}
\STATE{$M(k+1)=M(k)-\mu\frac{\Omega(k)\phi(k)\phi^T(k)}{\phi^T(k)\Omega(k)\phi(k)}M(k)+\frac{\phi(k)y(k+1)}{m^2(k)}$}
\ENDIF
\STATE{\textbf{Step 3}: Set $k+1\leftarrow k$ and \textbf{Step 2}}}}
\end{algorithmic}
\end{algorithm}
\begin{lem}\label{lemma:DF}
Suppose the augmented regressor matrix is obtained using Algorithm \ref{alg:DF-regressor}. In addition, we assume that the FE condition (see Definition 2) for the regressor vector (i.e. $\Phi(k)\in\mathbb{R}^{n\times 1}$) is satisfied. Then, there exist constants $\alpha_{\Omega} > 0$ and $\beta_{\Omega} > 0$ such that the following inequality holds:
\begin{align}\label{ma:Upper_Omega}
\alpha_{\Omega} I \leq \Omega(k)\leq \beta_{\Omega} I,\quad k\geq k_e,
\end{align}
where $k_e\in\mathbb{N}^{+}$ denotes the time step at which $\Omega$ satisfies the positive definiteness for the first time.
\end{lem}
\begin{proof}
See Theorems 1 and 2 in (\cite{DF}). Note that we were slightly modified.
\end{proof}

From Lemma \ref{lemma:DF}, by modifying Definition 1, we obtain the following.
\begin{thm}\label{thm:PE}
Assume that the regressor vector satisfies the FE condition.
Then, for the augmented regressor matrix $\Omega(k)$, $\delta = k_e >0$, $\alpha>0$, and $\beta>0$ exist such that
\begin{align}\label{ma:PE_Omega}
\alpha I\leq \sum_{k=\sigma}^{\sigma+\delta}\Omega^2(k)\leq \beta I,\ \forall \sigma\geq 0.
\end{align}
Therefore, the augmented regressor matrix satisfies the PE condition.
\end{thm}
\begin{proof}
From Lemma \ref{lemma:DF}, after time step $k_e$, the augmented regressor matrix $\Omega(k)$ maintains positive definiteness such that the longest interval width for the existence of the lower bound $\alpha>0$ is positive as $\delta=k_e$.
The upper bound, on the other hand, is obtained immediately from Lemma \ref{lemma:DF}.
\end{proof}

\begin{rem}
Using Algorithm \ref{alg:DF-regressor} and Lemma \ref{lemma:DF}, if the FE condition is satisfied for the regressor vector, the PE condition is satisfied for the augmented regressor matrix. This advantage is so meaningful that the parameter update using the augmented regressor can be applied to numerous control systems that require PE conditions.
\end{rem}
\begin{rem}
If the EF algorithm is applied to update the augmented regressor matrix, then the existence of the lower bound $\alpha$ of the matrix cannot be guaranteed without the PE condition for the regressor vector.
Therefore, the PE condition for the augmented regressor matrix cannot be guaranteed. However, the proposed method is highly effective, because it does not depend on the PE condition of the regressor vector.
\end{rem}

\subsection{Proposed RLS algorithm}
Based on the augmented regressor matrix in the previous section, we consider the following cost function minimization problem with respect to the estimated parameter $\hat{\theta}(k)$:
\begin{align}
J(k)&=\sum_{i=1}^{k}\lambda^{k-i}\left(\Omega(i)\hat{\theta}(k)-M(i)\right)^T\left(\Omega(i)\hat{\theta}(k)-M(i)\right)\notag\\
&\quad+\lambda^{k}(\hat{\theta}(k)-\theta(0))^TR(0)(\hat{\theta}(k)-\theta(0)),
\end{align}
where $\lambda\in(0,1)$ denote the value of forgetting factor, referred to as the outer forgetting factor in this paper.
Note that unlike inner forgetting, this corresponds to the same weight as the conventional forgetting factor.
Then, for all $k\geq 0$,
\begin{align}
\hat{\theta}(k+1)&=\hat{\theta}(k)-P(k)\Omega(k)N^{-1}(k)\left[\Omega(k)\hat{\theta}(k)-M(k)\right]\label{ma:law_theta}\\
P(k+1)&=\frac{1}{\lambda}\left(P(k)-\left(P(k)\Omega(k)N^{-1}(k)\Omega(k)P(k)\right)\right)\label{ma:law_P}
\end{align}
where $P(k)\in\mathbb{R}^{n\times n}$ and $R(k)\in\mathbb{R}^{n\times n}$ denote the covariance and information matrices, respectively.
Note that $R(k)\triangleq P^{-1}(k)$ and we set $P(0)=\gamma I,\ \gamma>0$.
Furthermore, $N(k)\triangleq \lambda I + \Omega(k)P(k)\Omega(k)$. Moreover, $\Omega(i)\hat{\theta}(k)-M(i)$ implies an augmented identification error, thus
\begin{align}
\Omega(k)\hat{\theta}(k)-M(k) = \Omega(k)\hat{\theta}(k)-\Omega(k)\theta=\Omega(k)\tilde{\theta}(k)
\end{align}
where $\tilde{\theta}(k)\triangleq \hat{\theta}(k)-\theta$,

\section{Stability analysis and discussion}
To discuss the stability, we first discuss the boundedness of the information matrix $R(k)$ as follows:
\begin{thm}\label{thm:P}
Consider Algorithm 1 and the parameter update law (\ref{ma:law_theta}).
Assume that the regressor vector $\phi(k)$ satisfies the FE condition.
For all $k\geq k_e$, an upper $\alpha\in\mathbb{R}>0$ and a lower $\beta\in\mathbb{R}>0$ bounds exist on the information matrix $R(k)$ such that
\begin{align}
\alpha I< R(k)< R(0)+\frac{1}{1-\lambda}\beta I
\end{align}
\end{thm}
\begin{proof}
This is based on the proof given by (\cite{EF-RLS}) and (\cite{Goel2020}).
The update law of the information matrix using the EF algorithm can be expressed as follows:
\begin{align}\label{ma:def_R}
R(k)=\lambda R(k-1)+\Omega^2(k-1)
\end{align}
Taking the sum of both sides in the interval $[\sigma,\sigma+\delta],\ \forall \sigma\geq 0$ and from Theorem \ref{thm:PE}, we obtain
\begin{align}\label{ma:R_temp}
\sum_{k=\sigma}^{\sigma + \delta}R(k)&=\sum_{k=\sigma}^{\sigma+\delta}(\lambda R(k-1)+\Omega^2(k-1))\nonumber\\
&\geq \sum_{k=\sigma}^{\sigma+\delta}\Omega^2(k-1)\geq \alpha I
\end{align}
From (\ref{ma:Upper_Omega}) and (\ref{ma:def_R}), it is clear that, because $R(k)> \lambda R(k-1)$ for all $\sigma\geq k_e$, (\ref{ma:R_temp}) can be expressed as
\begin{align}
&\sum_{i=0}^{\delta}\frac{1}{\lambda^{i}}R(\sigma+\delta)>\sum_{k=\sigma}^{\sigma+\delta}R(k)\geq \alpha I\\
&\iff R(\sigma+\delta)> \frac{\left(\frac{1}{\lambda}-1\right)}{\frac{1}{\lambda^{\delta +1}}-1}\alpha I\triangleq \alpha_R I>0
\end{align}
Next, we evaluate the upper bound of the information matrix $R(k)$.
From $\sum_{i=0}^{\delta}\lambda^iR(\sigma+1)<\sum_{k=\sigma}^{\sigma+\delta}R(k+1)$ for all $k\geq k_e$, we have
\begin{align}
R(\sigma+1)&<\frac{1-\lambda}{1-\lambda^{\delta+1}}\sum_{k=\sigma}^{\sigma+\delta}R(k+1)\\
&\leq\frac{1-\lambda}{1-\lambda^{\delta+1}}\left(\lambda^{\sigma}\sum_{k=0}^{\delta}R(k)+\frac{1-\lambda^{\sigma}}{1-\lambda}\beta I\right)\\
&\leq \lambda^{\sigma-\delta}R(\delta)+\frac{1-\lambda^{\sigma}}{1-\lambda^{\delta+1}}\beta I\\
&\leq R(\delta)+\frac{1}{1-\lambda^{\delta+1}}\beta I\triangleq \beta_R I>0
\end{align}
Therefore, $\alpha_R I<R(k)<\beta_R I,\ \forall k\geq k_e$.

After time step $k\geq k_e$, we can consider $\delta=0$.
This is because $\delta=k_e$ is the widest interval from Theorem \ref{thm:PE}, and the positive definiteness of the augmented regressor matrix is always satisfied after time steps $k_e$; therefore, we do not require the interval width.
Therefore, we have $\alpha_R I = \alpha I$ and $\beta_R I = R(\delta)+\frac{1}{1 - \lambda} \beta I$.
\end{proof}
\begin{rem}
Theorem \ref{thm:P} requires that the augmented regressor matrix $\Omega(k)$ satisfy the PE condition. If the EF algorithm as inner forgetting factor is applied to update the augmented regressor matrix without the PE condition for the regressor vector $\phi(k)$, then the existence of $\alpha>0$ satisfying (\ref{ma:PE_Omega}) in Theorem \ref{thm:PE} is not guaranteed.
In this case, the following theorem for Lyapunov analysis cannot guarantee stability.
On the other hands, the proposed method can guarantee Theorem \ref{thm:PE} without PE condition due to applying DF algorithm as inner forgetting factor.
\end{rem}

Next, using the boundedness results of the information matrix, the stability of the proposed method is proved via a Lyapunov stability analysis.
\begin{thm}\label{thm:ES}
Consider Algorithm 1 and the parameter update law (\ref{ma:law_theta}).
Assume that the regressor vector $\phi(k)$ satisfies the FE condition.
Subsequently, parameter error $\tilde{\theta}(k)$ globally exponentially converges to zero independent of the value of the EF forgetting factor $\lambda\in(0,1)$.
\end{thm}
\begin{proof}
Consider the following Lyapunov function candidate:
\begin{align}
V(k)=\tilde{\theta}^T(k)R(k)\tilde{\theta}(k)
\end{align}
From Theorem \ref{thm:P}, because the information matrix is bounded above and below, it follows that
\begin{align}
\alpha_R\|\tilde{\theta}(k)\|^2\leq V(k)\leq\beta_R\|\tilde{\theta}(k)\|^2
\end{align}
This holds true regardless of the PE condition, which is important in this study.
Next, as the covariance matrix $P(k),\ \forall k\geq 0$ is a positive definite matrix, the covariance matrix can be transformed as follows:
\begin{align}
&P(k+1)=\frac{1}{\lambda}\left(P(k)-\left(P(k)\Omega(k)N^{-1}(k)\Omega(k)P(k)\right)\right)\notag\\
&\iff \lambda P(k+1)P^{-1}(k)=I-P(k)\Omega(k)N^{-1}(k)\Omega(k)
\end{align}
Based on the above results, the parameter error equation can be expressed as follows:
\begin{align}
\tilde{\theta}(k+1)&=\tilde{\theta}(k)-P(k)\Omega(k)N^{-1}(k)\Omega(k)\tilde{\theta}(k)\notag\\
&=[I-P(k)\Omega(k)N^{-1}(k)\Omega(k)]\tilde{\theta}(k)\notag\\
&=\lambda P(k+1)P^{-1}(k)\tilde{\theta}(k)
\end{align}
Thus, evaluating the upper bound of the Lyapunov function yields:
\begin{align}
V(k+1)&=\lambda\tilde{\theta}^T(k+1)P^{-1}(k)\tilde{\theta}(k)\notag\\
&=\lambda\tilde{\theta}^T(k)[I-P(k)\Omega(k)N^{-1}(k)\Omega(k)]^TP^{-1}(k)\tilde{\theta}(k)\notag\\
&\leq \lambda V(k)
\end{align}
Therefore, the global exponential stability of the parameter error is guaranteed.
\end{proof}

\section{Simulation results}
In this section, we evaluate the effectiveness of the proposed method in two cases: (i) parameter convergence rate, and (ii) robustness against characteristic changes.
\subsection{Simulation condition}
Consider a mass-spring-damper system discretized with a sampling time $1$ s (\cite{MBD}).
In this study, three cases are also considered for mass $m=5$ kg: (a) $k=1$ N/m, $b=1$ Ns/m, (b) $k=10$ N/m, $b=0.01$ Ns/m, (c) $k=0.1$ N/m, $b=10$ Ns/m.
Thus, the ARX model can be expressed as follows:
\begin{align}
y(k+1) = \phi^T(k)\theta,
\end{align}
where $\phi(k)=[y(k),\ y(k-1),\ u(k),\ u(k-1)]^T$, and
\begin{itemize}
  \item[(a)] $\theta=[1.6405,\ -0.8187,\ 0.4606,\ 0.4307]^T$,
  \item[(b)] $\theta=[0.3116,\ -0.9980,\ 0.4218,\ 0.4215]^T$,
  \item[(c)] $\theta=[1.1267,\ -0.1353,\ 0.2834,\ 0.1482]^T$.
\end{itemize}
The initial values of the estimated parameters were set to $\hat{\theta}(0) = [0,\ 0,\ 0,\ 0]^T$, and the initial covariance matrix was set to $P(0) = 1000I$, that is, the initial information matrix was $R(0) = 10^{-3}I$.
Furthermore, $u(k)=\sin(0.1k)$ was selected as the identification input, which does not satisfy the PE condition, but satisfies FE condition.

\noindent \textbf{\underline{Case (i)}} This compares the conventional EF-RLS algorithm (\cite{EF-RLS}), CL with data collection algorithm (\cite{DCL}), and DF-based CL (\cite{DF-CL}) algorithms with the proposed method to evaluate the parameter convergence rate using only the system in (a).
The values of each forgetting factor were set as follows: $\lambda=0.99$ for EF-RLS algorithm, $\mu=0.5$ for DF-CL algorithm, and for the proposed method, $\lambda=0.99$ and $0.01$ for outer forgetting, whereas $\mu=0.99$ was commonly set for inner forgetting.

\noindent\textbf{\underline{Case (ii)}} This compares the conventional EF-RLS algorithm and the proposed method and its method without the inner forgetting factor or EF algorithm instead of the DF algorithm to evaluate its robustness to characteristic changes.
System generated characteristic changes of $0\leq k<200$ steps in (a), $200\leq k < 1200$ steps in (b), and $k>1200$ steps in (c).
The values of each forgetting factor were set as follows: $\lambda=0.99$ for the EF-RLS algorithm, $\mu=0.99$ for inner forgetting using the EF algorithm of the proposed method, $\mu=0.99$ and $0.01$ for inner forgetting using the DF algorithm of the proposed method, and $\lambda=0.01$ were commonly set for outer forgetting using the proposed method.

\subsection{Simulation results and discussion}
Figure 1 shows the minimum eigenvalue of a matrix based on regressor vector $\phi(k)\phi^T(k)$ and the augmented regressor matrix $\Omega^2(k)$ generated by Algorithm 1 using $\mu=0.99$ at each time step in Case (i).
The minimum eigenvalue of the matrix with regressor vector is almost zero, indicating that the PE condition is not satisfied. However, the minimum eigenvalue of the augmented regressor matrix used in the proposed method is always positive after time of $4$ steps, which corresponds to $k_e$.
This implies that the signal satisfies the PE condition at $\delta=4$, which is stronger than the PE condition because we can take $\delta=0$ after $k_e$ time steps.
Therefore, it can be seen that there is great value in generating the augmented regressor matrix of the proposed method.

Figure 2 shows the identification errors and Fig. 3 shows the parameter errors as logarithms on the vertical axis.
As shown in Fig. 2, all identification errors were close to zero, indicating accurate identification.
Figure 3 shows that the convergence of the parameters of the conventional EF-RLS algorithm stops at a constant value, clearly indicating that no true value convergence is observed. This is because it does not satisfy the PE condition, as illustrated in Fig. 1.
In contrast, it is known that the condition requiring the minimum eigenvalue of the cumulative information matrix to diverge, that is, $\sum_{k=0}^{\infty} \lambda_{\min}[\phi(k)\phi^T(k)] = \infty$, is a necessary and sufficient condition for guaranteeing the asymptotic stability of the parameter estimation error in the RLS algorithm without forgetting. Although this condition is weaker than the PE condition, it is difficult to satisfy in practice or in the present case.
The conventional CL achieves exponential convergence of the parameter error to $0$, but its convergence is very slow. Although this can be improved by incorporating the DF algorithm, the improvement remains limited compared with other methods. This is because the CL extends the gradient method, which theoretically constrains the gain of parameter updates to less than two.
However, the proposed method achieves fast parameter convergence because the adaptive gains in the RLS algorithm correspond to the covariance matrix $P(k)$ and can be updated with higher gains than the gradient methods.
Furthermore, by setting the forgetting factor $\lambda$ close to $0$, the designer can tune the parameter convergence rate in the design phase and reduce the parameter error to within the limit of the calculation accuracy.
In general, the RLS algorithm requires a conservative choice of the forgetting factor to avoid instability, and it is commonly recommended to set $\lambda$ between $0.99$ and $0.95$ (\cite{landou}). A significant contribution of the proposed method is that the forgetting factor can be arbitrarily selected, allowing designers to directly tune the convergence rate.
Hence, the proposed method can set $\lambda=0.01$, which is difficult for conventional RLS, and achieved a significant improvement in parameter convergence rate.

Figure 4 shows a comparison of the estimated parameters for characteristic changes.
From this figure, the proposed method based on DF with the inner forgetting set to $\mu=0.99$ achieves high convergence for all parameters and characteristic changes.
However, for $\mu=0.01$, the convergence rate is slower, although there is no oscillatory response owing to insufficient forgetting. This is evident from the minimum eigenvalues shown in Fig. 5 (a).
Both the EF-based proposed method and EF-RLS algorithm exhibited slow convergence. After a change in the system characteristics, the responses deviate from the true values or exhibit oscillations. This behavior can be attributed to the fact that the minimum eigenvalue of the covariance matrix, as shown in Fig. 5 (a), becomes small because the forgetting factor is set close to 1.
However, it may become unstable if the value of the inner forgetting factor obtained using the EF algorithm is set close to $0$. This is because the lower bound of the information matrix shown in Theorem \ref{thm:P} is not guaranteed and the adaptive gain becomes excessively high, making it sensitive to characteristic changes and noise.
The proposed method without inner forgetting did not converge with changes in the true value parameters. This does not improve the robustness to changes in characteristics even if outer forgetting is introduced, suggesting the need for the appropriate introduction of inner forgetting, which is one of the contributions of this study.
This is because, even if outer forgetting is introduced, previous data cannot be forgotten and are still more dominant in updating the augmented regressor matrix.
Finally, Fig. 5 (b) presents a comparison of the condition numbers in the covariance matrix.
This result also shows that the proposed method (with inner forgetting as DF) has the smallest value, indicating that it is highly robust against characteristic changes and noise.
This result indicates that the smallest eigenvalues in Fig. 5 (a) are sufficiently large and that it is important to select values for both inner $\mu$ and outer $\lambda$ forgetting that provide sufficiently high forgetting performance.

\begin{figure}[th]
\centering
\includegraphics[width=8.5cm]{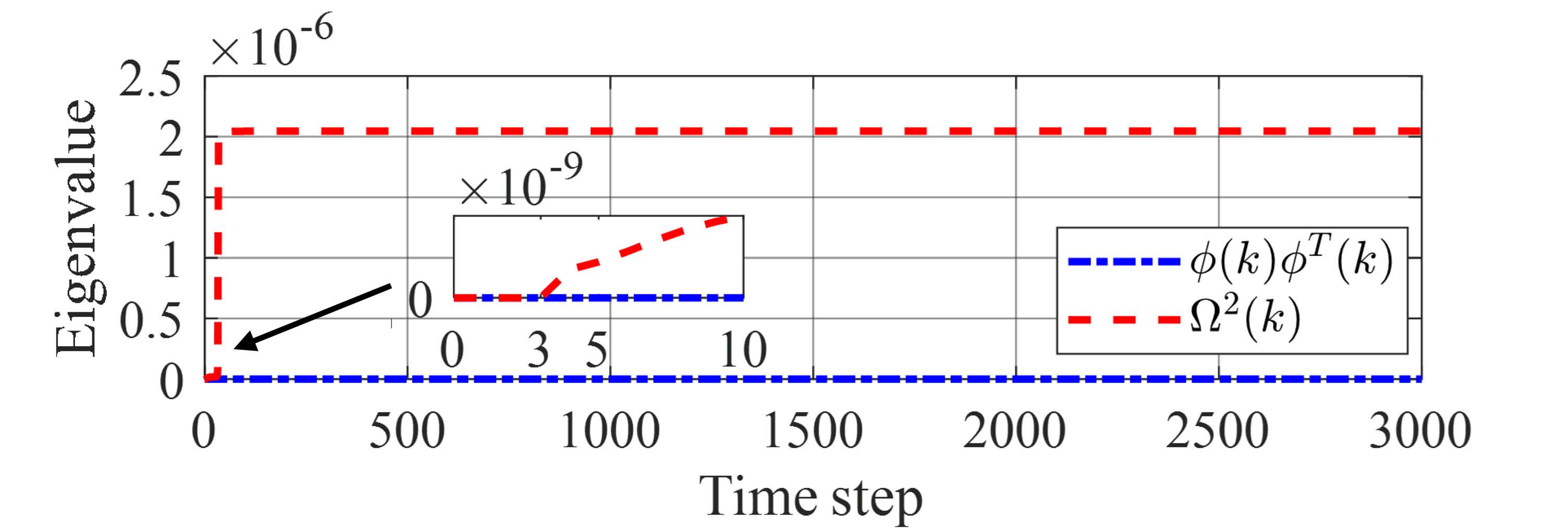}
  \caption{Comparison of minimum eigenvalues by regressor vector $\phi(k)\phi^T(k)$ and augmented regressor matrix $\Omega^2(k)$ in Case (i)}
  \label{fig:fig1}
\end{figure}

\begin{figure}[th]
\centering
\includegraphics[width=8.5cm]{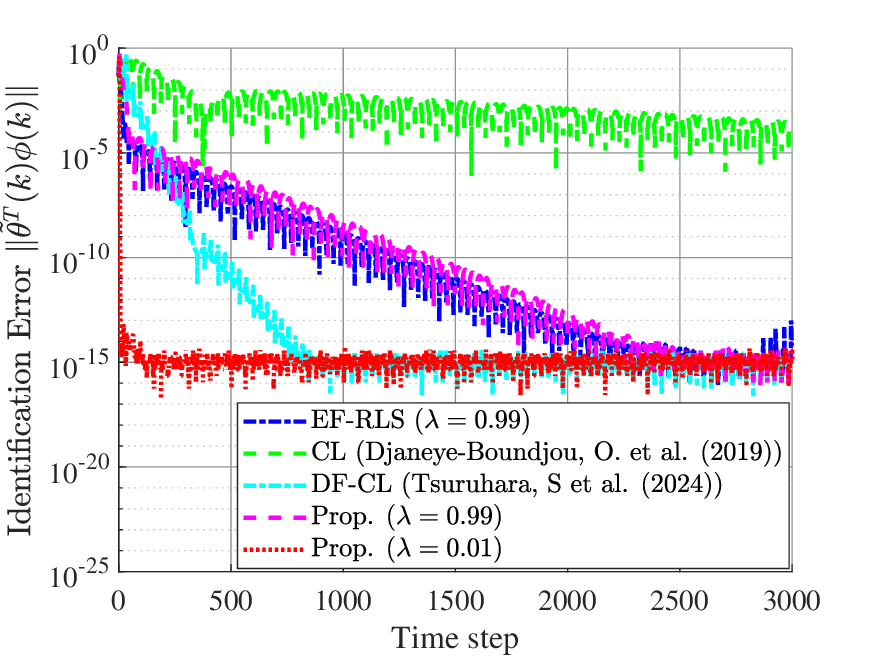}
  \caption{Comparison of identification errors in logarithmic scale under identification input $u(k)=\sin(0.1k)$ for Case (i) (EF-RLS: $\lambda=0.99$, DF-CL: $\mu=0.5$, Proposed method with inner forgetting factor: $\mu=0.99$). Note that the EF-RLS ($\lambda=0.01$) was destabilized.}
  \label{fig:fig2}
\end{figure}

\begin{figure}[th]
\centering
\includegraphics[width=8.5cm]{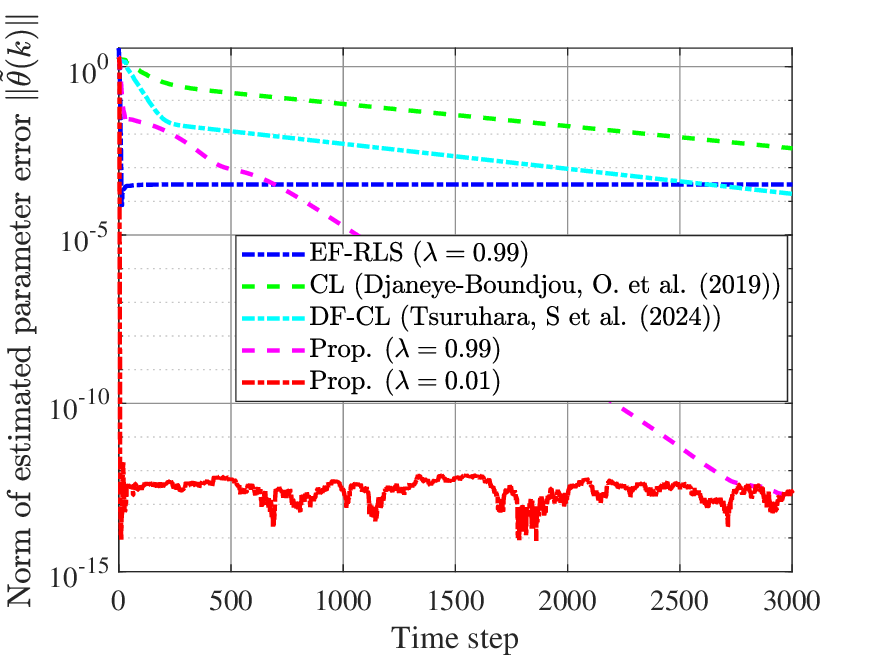}
  \caption{Comparison of parameter errors as the same condition in Fig. \ref{fig:fig2}.}
  \label{fig:fig3}
\end{figure}

\begin{figure}[th]
\centering
\includegraphics[width=8.5cm]{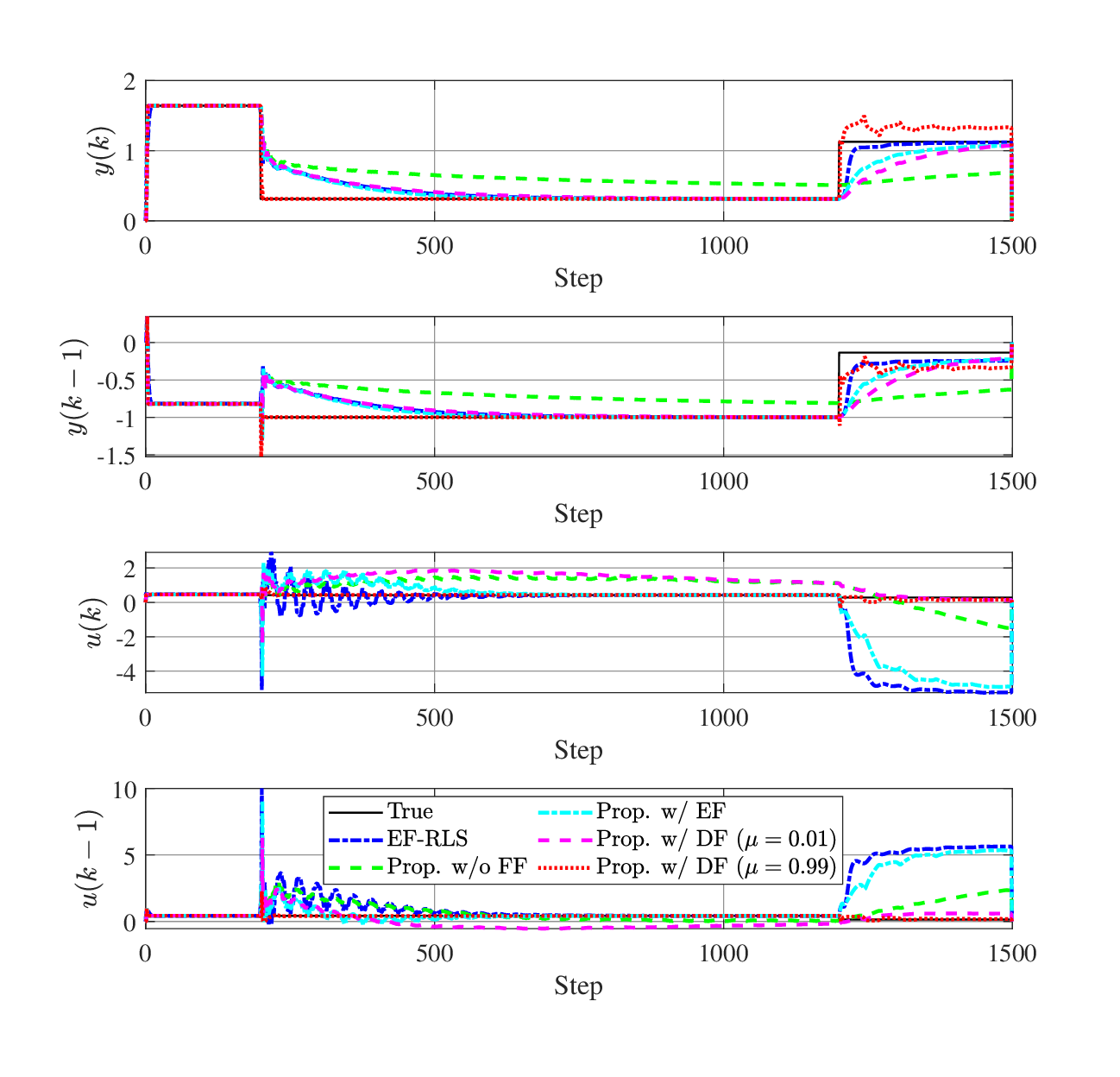}
  \caption{Comparison of estimated parameter under identification input $u(k) = \sin(0.1k)$ for Case (ii), which includes characteristics changes (EF-RLS: $\lambda = 0.99$, Proposed method with outer forgetting factor: $\lambda = 0.01$)}
  \label{fig:fig4}
\end{figure}

\begin{figure}[ht]
  \begin{minipage}[b]{0.49\linewidth}
    \centering
    \includegraphics[width=\linewidth]{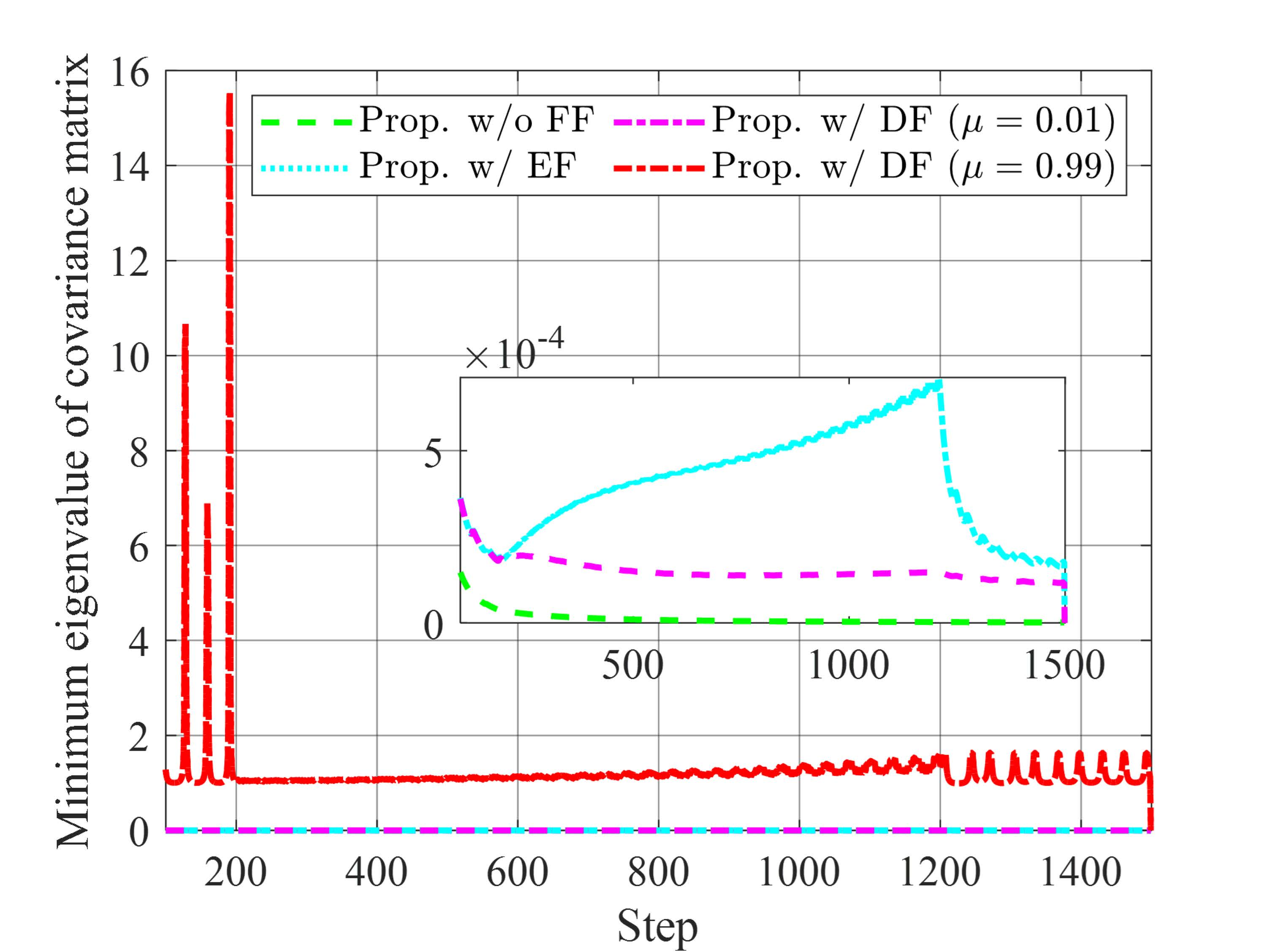}
    \subcaption{Minimum eigenvalue}
    \label{fig:fig5a}
  \end{minipage}
  \hfill
  \begin{minipage}[b]{0.49\linewidth}
    \centering
    \includegraphics[width=\linewidth]{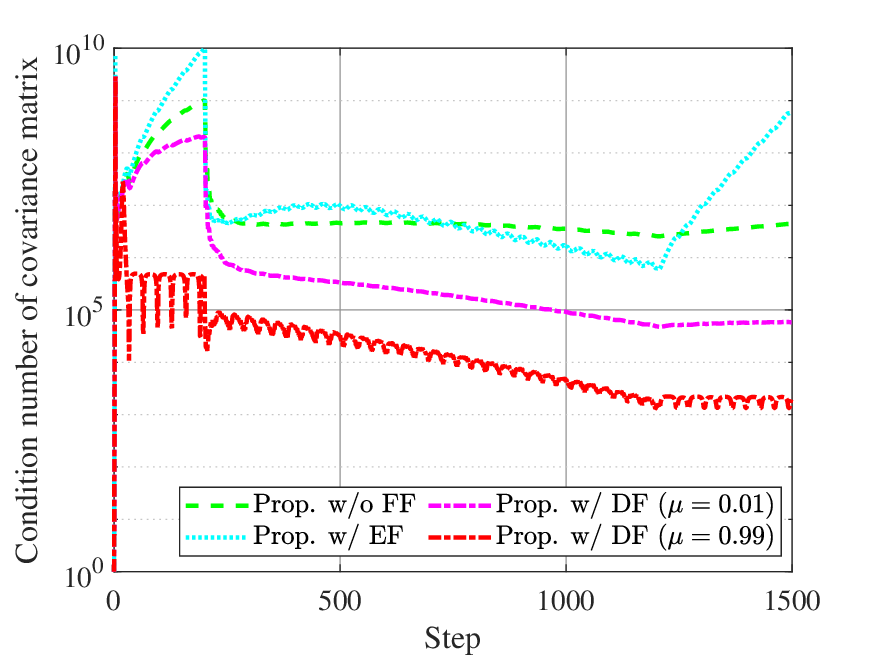}
    \subcaption{Conditon number}
    \label{fig:fig5b}
  \end{minipage}
  \caption{Comparison of minimum eigenvalue of covariance matrix as the same condition in Fig. \ref{fig:fig4}.}
\end{figure}
\section{CONCLUSIONS}
In this study, we proposed a novel discrete-time RLS algorithm with two layer forgetting factor that improves the fast true value convergence of the parameters and its robustness against characteristic changes under the FE condition of the regressor vector. It has been proven that the global exponential stability of the parameter errors and the design of convergence rates can be achieved without requiring PE conditions.
Numerical simulations illustrated the effectiveness of the algorithm compared with the conventional RLS algorithm. Several methods that do not require the PE property, particularly selecting arbitrary values of the forgetting factor, make a significant contribution compared with the conventional RLS algorithm.
Furthermore, the proposed method is expected to have a wide range of applications owing to its simple structure and no requirement of design parameters such as thresholds and/or weight parameters.
\bibliography{ifacconf}




\end{document}